\documentclass[prb,twocolumn,showpacs,preprintnumbers,amsmath,amssymb,superscriptaddress,10pt]{revtex4}

\usepackage{graphicx}
\usepackage{dcolumn}
\usepackage{bm}

\begin{document}

\title{
Epitaxial-strain effect on charge/orbital order in Pr$_{0.5}$Ca$_{0.5}$MnO$_3$ films}

\author{
D. Okuyama}%
\affiliation{
Cross-Correlated Materials Research Group (CMRG), ASI, RIKEN, Wako 351-0198, Japan}
\author{
M. Nakamura}%
\affiliation{
Cross-Correlated Materials Research Group (CMRG), ASI, RIKEN, Wako 351-0198, Japan}
\author{
Y. Wakabayashi}%
\affiliation{
Graduate School of Engineering Science, Osaka University, Toyonaka 560-8531, Japan}
\author{
H. Itoh}%
\affiliation{
Multiferroics Project, ERATO, Japan Science and Technology Agency (JST), 
c/o Department of Applied Physics, University of Tokyo, Tokyo 113-8656, Japan}
\author{
R. Kumai}%
\affiliation{
National Institute of Advanced Industrial Science and Technology (AIST), Tsukuba 305-8562, Japan}
\author{
H. Yamada}%
\affiliation{
National Institute of Advanced Industrial Science and Technology (AIST), Tsukuba 305-8562, Japan}
\author{
Y. Taguchi}%
\affiliation{
Cross-Correlated Materials Research Group (CMRG), ASI, RIKEN, Wako 351-0198, Japan}
\author{
T. Arima}%
\affiliation{
Institute of Multidisciplinary Research for Advanced Materials, Tohoku University, Sendai 980-8577, 
Japan}
\author{
M. Kawasaki}%
\affiliation{
Cross-Correlated Materials Research Group (CMRG), ASI, RIKEN, Wako 351-0198, Japan}
\affiliation{
WPI-AIMR, Tohoku University, Sendai 980-8577, Japan}
\author{
Y. Tokura}%
\affiliation{
Cross-Correlated Materials Research Group (CMRG), ASI, RIKEN, Wako 351-0198, Japan}
\affiliation{
Multiferroics Project, ERATO, Japan Science and Technology Agency (JST), 
c/o Department of Applied Physics, University of Tokyo, Tokyo 113-8656, Japan}
\affiliation{
National Institute of Advanced Industrial Science and Technology (AIST), Tsukuba 305-8562, Japan}
\affiliation{
Department of Applied Physics, University of Tokyo, Tokyo 113-8656, Japan}

\date{\today}

\begin{abstract}
Effect of growth orientation on charge- and orbital-ordering (CO-OO) phenomena has been studied for 
Pr$_{0.5}$Ca$_{0.5}$MnO$_3$ 
epitaxial thin films fabricated on (LaAlO$_3$)$_{0.3}$-(SrAl$_{0.5}$Ta$_{0.5}$O$_3$)$_{0.7}$ (LSAT) 
substrates by means of resistivity, synchrotron x-ray diffraction, and polarized optical microscopy measurements. 
CO-OO transition is observed around 220 K for a film grown on an LSAT (011) substrate ((011)-film), 
similarly to a bulk sample, while a film grown on a (001) plane of LSAT ((001)-film) shows 
much higher transition temperature around 300 K. 
The domain size of OO is approximately 3 times as large in the (011)-film as in the (001)-film. 
These results demonstrate that various properties of CO-OO phenomena can be controlled 
with the growth orientation via the epitaxial strain from the substrate. 
\end{abstract}

\maketitle

Manganites with perovskite-related structures have been extensively studied as functional materials\cite{Tokura2006}. 
Several colossal magnetoresistive manganites have been reported to exhibit gigantic responses to 
various kinds of stimuli, such as electric-field, x-ray, and light, in addition 
to magnetic field\cite{Asamitsu1997,Kiryukhin1997,Rini2007,Takubo2008}. 
In half-doped systems, charge- and orbital-ordering (CO-OO) instabilities often compete with, 
and dominate the ferromagnetic double exchange interaction, giving rise to insulating states. 
As a result, these materials are sensitive to perturbation, and the phase change accompanied by 
a huge decrease in resistivity is realized when subject to the external stimuli. 
Especially, that driven by voltage pulses\cite{Asamitsu1997} is of current interest related with the application 
in resistance-change nonvolatile memories\cite{Liu2000,Sawa2004}. 
As another method of phase control,  the growth orientation has recently proved to be effective 
in epitaxial films\cite{Nakamura2005}. 
A thermally induced transition from CO-OO insulator to ferromagnetic metal was observed in Nd$_{0.5}$Sr$_{0.5}$MnO$_3$ 
film on [011]-oriented SrTiO$_3$ substrate\cite{Nakamura2005,Wakabayashi2006}, 
similarly to a bulk sample\cite{Kuwahara1995}, 
whereas insulating behavior was observed for all the temperature range below room temperature 
in Nd$_{0.5}$Sr$_{0.5}$MnO$_3$ film on [001]-oriented SrTiO$_3$ substrate\cite{Nakamura2005}. 
In this study, we have fabricated half-doped manganite Pr$_{0.5}$Ca$_{0.5}$MnO$_3$ thin films 
on (LaAlO$_3$)$_{0.3}$-(SrAl$_{0.5}$Ta$_{0.5}$O$_3$)$_{0.7}$ (LSAT) substrates with different orientation, 
and performed electrical resistivity, synchrotron x-ray diffraction, and optical experiments. 
Pr$_{0.5}$Ca$_{0.5}$MnO$_3$ is a prototypical system, showing a clear CO-OO state 
which is sufficiently robust for a wide range of hole doping (0.3$\le$$x$$\le$0.7) with a high CO-OO transition 
temperature $T_{\mathrm{CO}}$=230 K\cite{Tomioka1996,Jirak2000,Zimmermann2001}. 
The averaged lattice constant of bulk Pr$_{0.5}$Ca$_{0.5}$MnO$_3$ is smaller than that 
of LSAT substrate by 1.4\%, resulting in the slight tensile strain normal to the growth orientation. 
We have found a significant change of the CO-OO states in terms of transition temperature and domain size 
due to the difference in the growth orientation via the tensile strain from the substrate. 

\begin{figure}
\includegraphics*[width=80mm,clip]{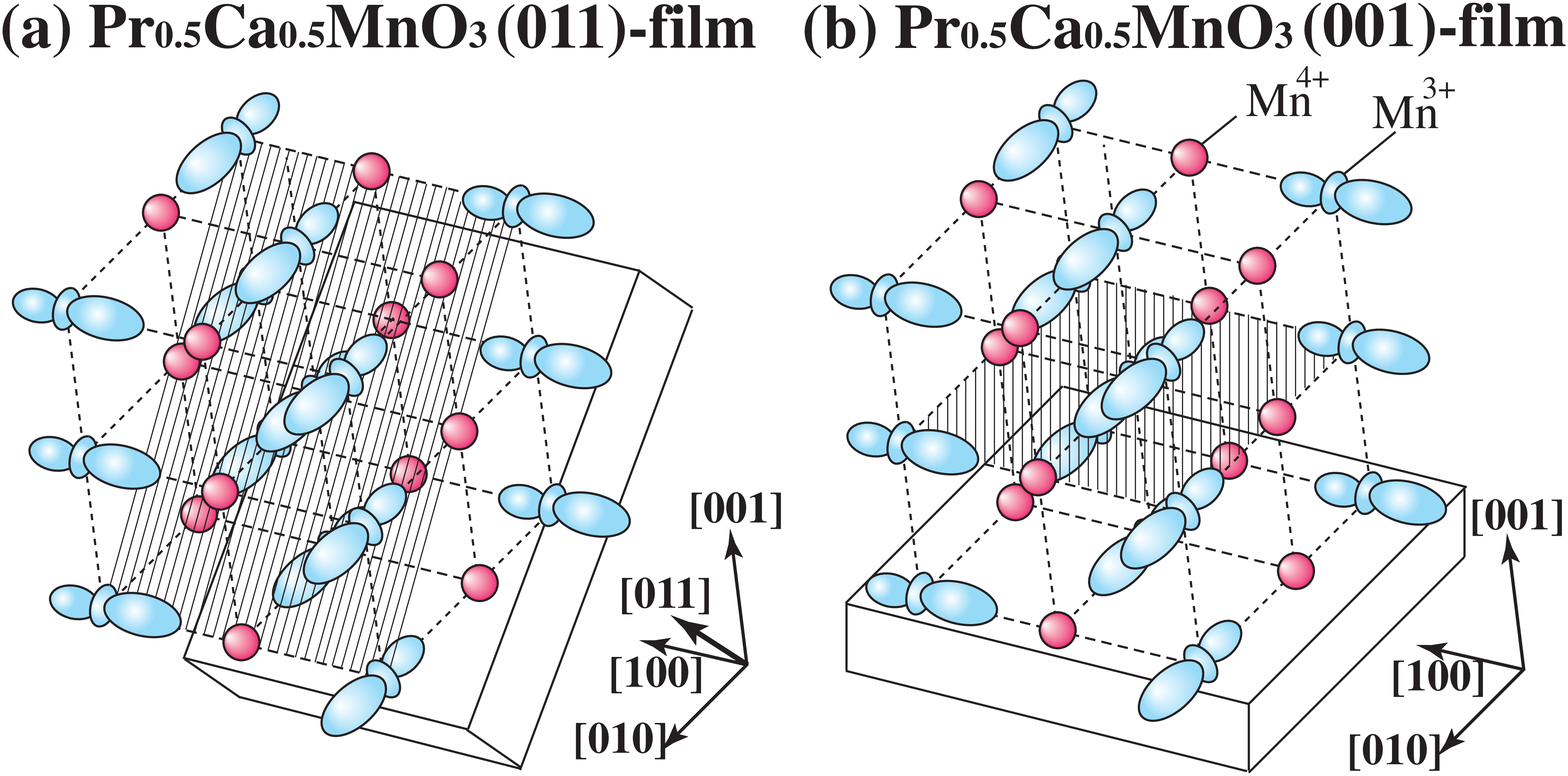}
\caption{\label{fig_01} 
(Color Online) 
Schematic view of charge- and orbital-ordered states in Pr$_{0.5}$Ca$_{0.5}$MnO$_3$ thin films on 
(a) [011]-oriented, and (b) [001]-oriented (LaAlO$_3$)$_{0.3}$-(SrAl$_{0.5}$Ta$_{0.5}$O$_3$)$_{0.7}$ 
(LSAT) substrate. 
Lobes represent occupied $e_{g}$ orbital at Mn$^{3+}$ sites aligned in a $(3x^2-r^2)$/$(3y^2-r^2)$ manner. 
Circles correspond to Mn$^{4+}$ sites. 
Hatched planes represent the planes parallel to the substrate-surface. 
The orientation of the crystal axes is indicated with the pseudo-cubic setting. 
}
\end{figure}

Pr$_{0.5}$Ca$_{0.5}$MnO$_3$ films on [011]- and [001]-oriented substrates (denoted as (011)-film and (001)-film, 
respectively, hereafter) were grown by pulsed laser deposition technique 
with a laser pulse frequency of 2 Hz at 850 $^{\circ}$C in an oxygen pressure of 1.5 mTorr. 
The total thickness of the films is approximately 400 \AA. 
X-ray diffraction experiments were performed on beamlines 3A and 4C at the Photon Factory, KEK, Japan. 
The photon energy of the incident x-rays was 9.5 keV. 
Optical birefringence signal was measured by using a polarized optical microscope equipped with a CCD detector. 
Estimated spatial resolution is less than 1 $\mu$m, i.e. around light wavelength. 
In Fig. \ref{fig_01}, we show schematic illustration on the relation between pseudo-cubic perovskite 
lattice and the LSAT substrates. 
We adopt the pseudo-cubic setting for denoting the crystal axes. 
Both of the $a$- and $b$-lattice constants are locked to those of substrate in the (001)-film, 
as evidenced by the x-ray diffraction pattern of (0 2 2) and (-1 0 3) reflections. 
By contrast, the $a$-lattice constant and [01-1]-axis are locked in the (011)-film, 
while $b$, $c$, and the angle between them have some freedom, 
as clearly revealed by an inspection of (0 2 1), (0 1 2), and (2 2 2) reflections. 

\begin{figure}
\includegraphics*[width=80mm,clip]{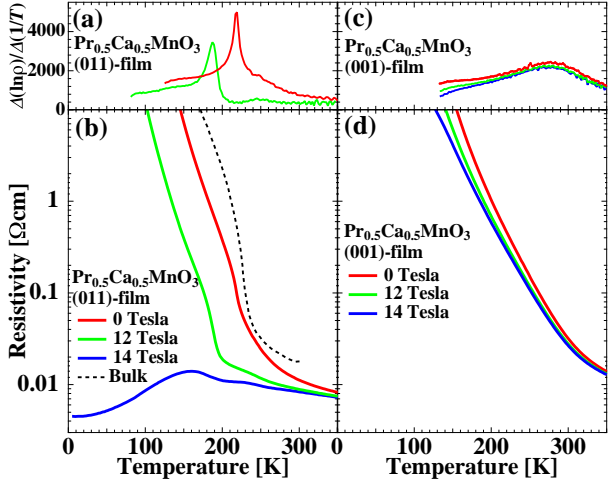}
\caption{\label{fig_02} 
(Color Online) 
Temperature dependence of resistivity in various magnetic fields for (b) the (011)-film and (d) the (001)-film. 
The $\varDelta(\ln{\rho})/\varDelta(1/T)$ calculated from the resistivity ($\rho$) data 
is plotted against temperature for (a) the (011)-film and (c) the (001)-film. 
Dashed line in (b) is the resistivity of a bulk sample in zero field reproduced from Ref. \onlinecite{Tomioka1996}. 
}
\end{figure}

Figure \ref{fig_02} shows the temperature dependence of resistivity in various magnetic fields. 
As shown in Fig. \ref{fig_02} (b), the (011)-film shows a clear anomaly associated with CO-OO transition 
around 220 K in zero filed, similarly to the bulk sample\cite{Tomioka1996}. 
The anomaly is manifested more clearly in the plot of $\varDelta(\ln{\rho})/\varDelta(1/T)$ in Fig. \ref{fig_02} (a). 
The application of 12 T reduces the CO-OO transition temperature by 30 K, but the transition still takes place. 
With further increasing the magnetic field, the CO-OO transition collapses at 14 T, and instead a metallic state 
shows up at low temperatures. 
By contrast, the (001)-film does not show such a clear anomaly as observed in the (011)-film, as plotted 
in Fig. \ref{fig_02} (d). 
However, a very broad peak structure is observed around 280 K (Fig. \ref{fig_02} (c))
in $\varDelta(\ln{\rho})/\varDelta(1/T)$, 
which turns out to almost coincide with the CO-OO transition temperature, as evidenced later by x-ray diffraction 
and optical birefringence (\textit{vide infra}). 
This temperature is clearly higher than that of the bulk sample, in accord with the results 
in Refs. \onlinecite{Ogimoto2005} and \onlinecite{Yang2006}. 
In addition, the CO-OO is robust against the magnetic field at least below 14 T. 
These transport data imply that the CO-OO state is stabilized for the (001)-film in comparison with the (011)-film. 

\begin{figure} 
\includegraphics*[width=80mm,clip]{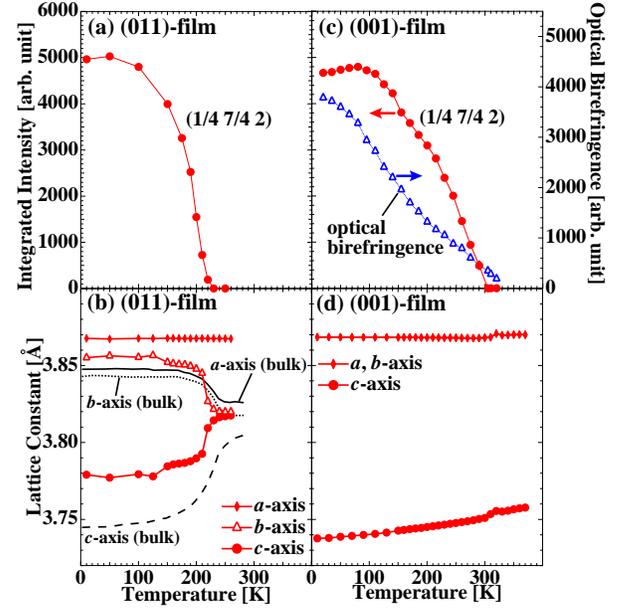}
\caption{\label{fig_03} 
(Color Online) 
Temperature dependence of integrated intensity of (1/4 7/4 2) superlattice reflection, which corresponds to the 
orbital-ordering, for (a) the (011)-film and (c) the (001)-film. 
Lattice constants are plotted against temperature for (b) the (011)-film and (d) the (001)-film. 
The lattice constants of the bulk sample\cite{Jirak2000} are also plotted in (b). 
In the panel (c), the intensity of optical birefringence is also shown. 
}
\end{figure}

In order to clarify the growth orientation dependence of the CO-OO transition in more detail, 
we performed a synchrotron x-ray diffraction experiment. 
Figure \ref{fig_03} (a) shows the integrated intensity of (1/4 7/4 2) superlattice reflection, 
which corresponds to the characteristic Jahn-Teller distortion with (1/4 1/4 0) propagation vector, for the (011)-film. 
The superlattice reflection begins to grow at 220 K in accord with the anomaly in resistivity. 
In Fig. \ref{fig_03} (b), lattice constants of the (011)-film are plotted\cite{footnote_01} along with those of the bulk sample. 
Upon the CO-OO transition, the $b$- and $c$-axes are elongated and shortened, respectively. 
This behavior is similar to the bulk sample as indicated by dotted and dashed lines, 
and is allowed by the absence of the locking of the $b$- or $c$-axis to the substrate. 
The temperature dependence of the superlattice reflection and lattice constants of the (001)-film 
is different from that of the (011)-film and/or bulk, as shown in Figs. \ref{fig_03} (c) and (d). 
The integrated intensity of the (1/4 7/4 2) superlattice reflection subsists at higher temperatures than $T_{\mathrm{CO}}$ 
of the (011)-film and the bulk sample, and finally disappears at $T_{\mathrm{CO}}^{(001)}\sim$300 K, 
in accord with the resistivity data and the result reported in Ref. \onlinecite{Yang2006}. 
While the $c$-lattice constant calculated from the (0 0 3) Bragg reflection of the film shows a tiny anomaly 
at $T_{\mathrm{CO}}^{(001)}$, the $a$- and $b$-lattice constants, which are locked by the substrate, 
show little temperature dependence, making marked contrast with the (011)-film. 
Due to the locking of the $a$- and $b$-lattice constants, the change in the $c$-lattice constant is also suppressed. 
The locked $a$- and $b$-lattice constants are sufficiently larger than that of the $c$-axis, 
similarly to the bulk sample in the ordered phase. 
This would be the origin of the stabilization of CO-OO states in the (001)-film. 
On the other hand, the $b$- and $c$-lattice parameters are not locked in the (011)-film, 
and therefore the observed behavior rather resembles that of the bulk sample. 

\begin{figure}
\includegraphics*[width=80mm,clip]{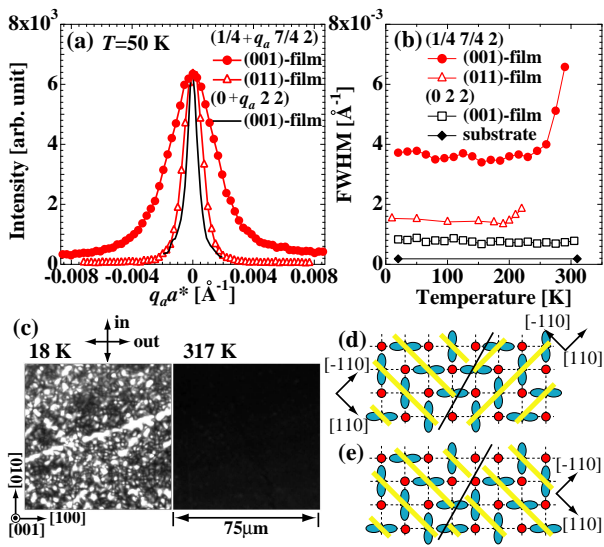}
\caption{\label{fig_04} 
(Color Online) 
(a) X-ray diffraction profiles of (1/4 7/4 2) superlattice reflection along the $a^{*}$-axis for 
the (001)- and the (011)-films at 50 K. 
The profile of (0 2 2) fundamental reflection for the (001)-film is also shown. 
(b) Temperature dependence of full width at half maximum (FWHM) of (1/4 7/4 2) reflections. 
We also show the FWHM of (0 2 2) reflection of the (001)-film and LSAT-substrate. 
(c) Polarized optical microscope images of the (001)-film at 18 K and 317 K, which are below and 
above the CO-OO transition temperature, respectively. 
Crossed arrows show the polarization directions of the incident (in) and transmitted (out) lights. 
(d) Boundary between two domains with different directions of orbital stripe. 
Orbital stripes are indicated by diagonal lines, which are parallel to the [110]-axis. 
(e) Schematic view of an anti-phase-type domain boundary. 
Across the domain wall, the phase of the charge- and orbital-density wave changes, 
while the optical anisotropy is kept unchanged. 
}
\end{figure}

In addition to the temperature dependence, the size of the ordered domains gives further information 
on the difference in growth orientation. 
Figure \ref{fig_04} (a) shows the profiles of (1/4 7/4 2) superlattice reflection and (0 2 2) fundamental lattice reflection 
of the thin films at 50 K. 
The (1/4 7/4 2) superlattice profile of the (001)-film is substantially broader than that of the (011)-film, which is 
as narrow as the fundamental lattice reflection. 
From these profiles, full width at half maximum (FWHM) is obtained, allowing the estimation of domain size. 
Thus obtained size of the orbital-ordered domain along the $a$-direction at 50 K is 
about 370 \AA \ and 1000 \AA \ for the (001)-film and the (011)-film, respectively, 
while that of crystal lattice is 1900 \AA. 
Note that the instrumental resolution limit is greater than 7000 \AA. 
The FWHM is almost temperature independent at low temperatures, 
but shows a rapid broadening towards the respective $T_{\mathrm{CO}}$ for both films, as shown in Fig. \ref{fig_04} (b).

The short-ranged nature of the CO-OO state in the (001)-film is also confirmed by 
the optical birefringence signal. 
Figure \ref{fig_04} (c) represents polarized optical microscope images of the (001)-film with the crossed Nicols configuration. 
At 317 K ($T> T_{\mathrm{CO}}^{(001)}$), no optical anisotropy is indicated by the completely dark image, 
but the bright image at 18 K clearly shows the emergence of the anisotropy. 
In the CO-OO state, the optical-principal axes coincide with the [110]- and [-110]-axes, 
giving rise to the anisotropy between the orbital stripe direction (indicated by diagonal lines in Figs. \ref{fig_04} (d) and (e)) 
and the orbital zig-zag chain direction\cite{Ishikawa1999}. 
(The temperature dependence of the integrated intensity of the birefringence is in accord with that of an x-ray 
superlattice reflection, as shown in Fig. \ref{fig_03} (c).) 
Moreover, the brightness of the image at 18 K is not uniform, indicating that the distribution of the domain size 
depends on the position. 
The birefringence signal is strong in intensity at the positions where the domain size is sufficiently larger than 
the wavelength of the light ($\sim$ several thousand \AA). 
On the other hand, the intensity should be reduced at the position where domains with different directions 
of orbital stripes (Fig. \ref{fig_04} (d)), i.e. different directions of fast- and slow-axes, 
assemble within the range of wavelength (or spatial resolution), 
because the spatially-averaged refractive index is less anisotropic. 
Therefore, the bright but non-uniform image at 18 K implies that the typical domain size is less than several thousand \AA, 
although a more quantitative estimate from the optical data is difficult. 

There are two factors that characterize the CO-OO domain. 
One is the direction of the orbital stripe as depicted in Fig. \ref{fig_04} (d), 
and the other is the phase of charge- and orbital-density wave in Fig. \ref{fig_04} (e). 
In the case of the (001)-film, fine structures of domains with different direction of orbital stripe would form so as to 
recover the global tetragonality that is imposed by the locking of the $a$- and $b$-lattice parameters to the substrate. 
On the other hand, this constraint is weakened in the case of the (011)-film due to the unlocked $b$-lattice constant. 
This would be the main origin of smaller domain size of the (001)-film than that of the (011)-film. 

In summary, we have performed resistivity, synchrotron x-ray diffraction, and polarized microscopy experiments 
for Pr$_{0.5}$Ca$_{0.5}$MnO$_3$ thin films fabricated on [001]- and [011]-oriented LSAT substrates. 
We have found that the CO-OO transition temperature of the (001)-film is much higher than that of the (011)-film 
and bulk sample. 
In addition, the domain size of orbital order is smaller in the (001)-film than in the (011)-film. 
These results demonstrate the controllability of the various properties of the CO-OO states in half-doped manganites 
with the change of the growth orientation via the epitaxial strain from the substrate. 

This study was performed with the approval of the Photon Factory Program Advisory Committee (No.2006S2-005) 
and (No.2008S2-004).


\begin{thebibliography}{200}
\bibitem{footnote_01}
The $a$-lattice parameter is estimated from that of substrate, and 
the $b$- and $c$-lattice constants are estimated from (0 2 1), (0 1 2), and (0 2 2) Bragg reflections of the film. 
The angle between $b$- and $c$-axes is evaluated to be 88.5$^{\circ}$ at 10 K.
\end{thebibliography}
\end{document}